\documentclass[superscriptaddress, reprint, amsmath,amssymb, aps,]{revtex4-2}

\usepackage{color}
\usepackage{graphicx}
\usepackage{dcolumn}
\usepackage{bm}

\usepackage{pgf}
\usepackage{tikz}
\usepackage{multirow}

\usetikzlibrary{arrows,decorations.pathmorphing,backgrounds,positioning,fit,petri}
\usetikzlibrary{decorations.pathreplacing}
\usetikzlibrary{decorations.markings}
\usetikzlibrary{decorations.shapes}
\usetikzlibrary{arrows.meta}
\usetikzlibrary{quotes,angles}
\usetikzlibrary{positioning}
\usetikzlibrary{patterns} 
\usetikzlibrary{chains}
\usetikzlibrary{hobby}

\begin{document}


\title{Two-point functions  in boundary loop models}

\author{Max Downing}
\email{max.downing@phys.ens.fr}
\affiliation{Laboratoire de Physique de l’École Normale Supérieure, ENS, Université PSL,
CNRS, Sorbonne Université, Université Paris Cité, Paris, France}

\author{Jesper Lykke Jacobsen}
\email{jesper.jacobsen@ens.fr}
\affiliation{Laboratoire de Physique de l’École Normale Supérieure, ENS, Université PSL,
CNRS, Sorbonne Université, Université Paris Cité, Paris, France}
\affiliation{Institut de physique théorique, CEA, CNRS, Université Paris-Saclay}

\author{Rongvoram Nivesvivat}
\email{rongvoramnivesvivat@gmail.com}
\affiliation{New York University Abu Dhabi, Abu Dhabi, United Arab Emirates}

\author{Hubert Saleur}
\email{hubert.saleur@ipht.fr}
\affiliation{Institut de physique théorique, CEA, CNRS, Université Paris-Saclay}
\affiliation{ Department of Physics and Astronomy, University of Southern California, Los Angeles}

\date{\today}

\begin{abstract}
Using techniques of conformal bootstrap, we propose analytical expressions for a large class of two-point functions of bulk fields in critical loop models defined on the upper-half plane. Our results include the two-point connectivities in the Fortuin--Kasteleyn random cluster model with both free and wired boundary conditions. We link the continuum expressions  to lattice quantities by computing  universal ratios of amplitudes for the two-point connectivities, and find excellent agreement with transfer-matrix numerics.
\end{abstract}

\maketitle


\paragraph*{\bf Motivation and goals.} 

Loop models ---  random ensembles of non-intersecting loops on a  two-dimensional lattice --- encompass several well-known statistical models, such as percolation, self-avoiding walks, the Fortuin--Kasteleyn (FK) random cluster model \cite{fk72}, and the $O(n)$ loop model \cite{nie82}. The continuum limit of these models at criticality, henceforth called {\sl critical loop models}, exhibit conformal invariance and can be studied by techniques of two-dimensional conformal field theories (CFT). Due to their intrinsic non-unitarity and non-rationality, these are difficult models. After some quick initial progress on the determination of critical exponents, both with physics techniques \cite{nie82,S86,dup88}, then later by the SLE formalism \cite{rs04,b08}, the field was on hold for many years. The bootstrap approach finally led to the (almost complete) solution \cite{jrs22,nrj23} of the problem in the bulk, with the determination  of four-point functions and operator product expansions (OPE) (up to the factorisation of four-point structure constants into OPE coefficients). Here again, a close relationship with results by probabilists emerged, incarnated in studies of conformal loop ensembles \cite{acsw21, jnrr25}.

For rational CFTs, the solution of the bulk and boundary problems are closely related, while for loop models the situation is very different. In the boundary case, some results for boundary critical exponents have been known for many years \cite{ds86},  but correlation functions of most interesting operators have resisted determination up to now. An exception to this is the case of degenerate fields \cite{car04}, which occur e.g.  in Cardy's celebrated crossing probability \cite{car91}, which was later proven rigorously by Smirnov \cite{smir01}. In this Letter, we introduce a general method  to compute two-point functions of any field in critical loop models defined on the upper-half plane. We propose analytical expressions for a large class of those two-point functions, and we interpret them physically as probabilities.




\paragraph*{\bf Critical loop models.} 
We frame our results in terms of a completely-packed loop model on the square lattice, defined on the upper-half plane. Starting from a $Q$-state Potts model with nearest-neighbor interaction energy $-J \delta_{\sigma_i,\sigma_j}$ along each lattice edge $(ij) \in E$, usual FK expansion \cite{fk72} gives the partition function
\begin{equation}
 \label{FKexpansion}
 Z_{\rm FK} = \sum_{A \subseteq E} v^{|A|} Q^{K_0} Q_1^{K_1} \,,
\end{equation}
where $v = {\rm e}^J - 1$ has the critical value $v_{\rm c} = \sqrt{Q}$ by a duality argument.
When $v \uparrow v_{\rm c}$, \eqref{FKexpansion} undergoes a second-order phase transition, the size $|A|$ of the largest cluster (connected component) in $A$ diverges, and the continuum limit becomes conformally invariant.
A weight $Q$ (resp.\ $Q_1$) is given to each of the $K_0$ (resp.\ $K_1$) clusters that contains none (resp.\ at least one) of the boundary sites, and we take $Q_1 = Q$ (resp.\ $Q_1 = 1$) for free (resp.\ fixed) boundary conditions (BC).
On the medial lattice (whose vertices are the mid points of $E$), draw the contours separating the clusters of $A$ from the dual clusters of $E \setminus A$. Each such contour is called a {\em loop}. One finds \cite{bkw76} that
$Z_{\rm FK} \propto \sum_{\text{loops}} Q^{\ell_0/2} Q_1^{\ell_1/2}$, meaning that a weight $\sqrt{Q}$ (resp.\ $\sqrt{Q_1}$) is given to each of the $\ell_0$ (resp.\ $\ell_1$) loops that does not (resp.\ does) touch the boundary.
 
Using this {\em loop representation}, the Coulomb Gas (CG) formalism of \cite{nie82} maps \eqref{FKexpansion} to a solid-on-solid model, whose critical limit is a Gaussian free field modified by background and screening charges. The critical loop model is hence described by a CFT whose central charge $c$ is related to $Q$ by the CG coupling $\beta^2$:
\begin{equation}
Q = 4\cos^2(\pi\beta^2)
\quad\text{and}\quad
c = 1-6(\beta - \beta^{-1})^2
\ .
\end{equation}
The list of local operators in this CFT can be read off from the torus partition function of \cite{fsz87} and the annulus partition function of \cite{djs09}. We write $\Psi(z)$\footnote{While  standard in papers about boundary CFT, we emphasize this notation does not imply the field is an analytic function of $z$.}  for an operator acting at $z = x + i y$, and denote $(\Delta,\bar{\Delta})$ its conformal weights, i.e.\ the eigenvalues under dilations of $z$ and the mirror image $\bar{z} = x - i y$. For $N \in \mathbb{N}_{>0}$, the primary operators of relevance to this Letter are:
\begin{itemize}
\item $V^d_{(1, N)}(z)$: Level-$N$ degenerate energy-type operator of weights $(\Delta_{(1,N)}, \Delta_{(1,N)})$;
\item $V_{\left (N/2,0\right)}(z)$:  Bulk $N$-leg (or fuseau, or watermelon) operator which inserts $N$ open loop segments at $z$;
\item $V_{\left (0,1/2 \right)}(z)$: Spin operator which inserts an FK cluster at $z$;
\item $v_{(N,1)}(x)$: Boundary $(N-1)$-leg operator which inserts $N-1$ open loop segments at $x$ (the  real axis).
\end{itemize}
 We write $V_{(r, s)}$ for the bulk primary of weights $(\Delta_{(r,s)}, \Delta_{(r,-s)})$, and $v_{(r,s)}$ for the boundary primary of weight $\Delta_{(r,s)}$. We use the Kac parametrization $\Delta_{(r,s)} = P_{(r,s)}^2 -P_{(1,1)}^2$, where
$P_{(r,s)} = \frac12\left(r\beta -s \beta^{-1}\right)$.


\paragraph*{\bf Conformal bootstrap.}

Conformal invariance requires  the  upper-half plane two-point function of bulk fields to take the form \cite{car04}:
\begin{multline}
\langle V_i(z_1) V_j (z_2) \rangle 
=\frac{G_{ij}(\sigma)}{ |z_1 - \bar z_2|^{4\Delta_{i}}(z_2 - \bar z_2)^{2\Delta_{j} -2\Delta_{i}}} \,,
\label{12}
\end{multline}
where the cross-ratio is $\sigma = \frac{|z_1 - z_2|^2}{(z_1 -\bar z_2)(\bar z_1 - z_2)}$, and we restrict in what follows to $i,j  \in \{ \left(N/2, 0\right), \left(0, 1/2 \right)\}$.

The bootstrap approach relies on the associativity of the OPE \cite{bpz84}, which translates  into crossing-symmetry for \eqref{12}. This means that \eqref{12} must be consistent under either approaching the two bulk fields to one another ({\em bulk channel}, or $s$-channel expansion: $\sigma \to 0$), or approaching each field to the boundary ({\em boundary channel}, or $t$-channel expansion: $\sigma \to 1$). The consistency can be represented as the  standard identity \cite{cl91, lew92}
\[
\sum_{\Delta \in \mathcal{S}^{(s)}} 
D^{\text{bulk}}_\Delta
\begin{tikzpicture}[baseline=-0.5ex, scale=1]
  \coordinate (v) at (0,0);
  \coordinate (l) at (-0.6,0.6);
  \coordinate (r) at (0.6,0.6);
  \coordinate (d) at (0,-0.6);

  \draw (v) -- (l);
  \draw (v) -- (r);
  \draw (v) -- (d);
  \draw (-0.8,-0.6) -- (0.8,-0.6);

  \node at (-0.6,0.8) {$\Delta_i$};
 \node at (.7,0.8) {$\Delta_j$};

  \node at (-0.2,-0.25) {$\Delta$};
\end{tikzpicture}
\;=\;
\sum_{\Delta \in \mathcal{S}^{(t)}}
D^{\text{bdy}}_\Delta
\begin{tikzpicture}[baseline=-0.5ex, scale=1]
  \coordinate (l1) at (-0.6,0.6);
  \coordinate (l2) at (0.6,0.6);
  \coordinate (d1) at (-0.6,-0.6);
  \coordinate (d2) at (0.6,-0.6);

  \draw (l1) -- (d1);
  \draw (l2) -- (d2);

  \draw (-0.8,-0.6) -- (0.8,-0.6);

  \node at (-0.6,0.8) {$\Delta_i$};
 \node at (.6,0.8) {$\Delta_j$};
  \node at (0,-0.85) {$\Delta$};
\end{tikzpicture} \,,
\]

\noindent where the horizontal line  represents the boundary $y=0$. Introducing mirror images of the fields \cite{Cardy84} gives the equivalent identity

\begin{align}
  \sum_{\Delta \in \mathcal{S}^{(s)}} 
 D^{\text{bulk}}_\Delta
 \hspace{-0.8cm}
 \begin{tikzpicture}[baseline=(base), scale = .28]
\coordinate (base) at (0, 2);
\draw (2, -1) -- (0,0) -- (0, 4) -- (2,5);
\draw (-2, -1) -- (0,0);
\draw (0,4) -- (-2,5);
\node [left] at (2,2){$\Delta$};
\node at (2.5, -1.7) {$\Delta_{j}$};
\node at (2.5, 5.7) {$\Delta_{j}$};
\node at (-2.5, -1.7) {$\Delta_{i}$};
\node at (-2.5, 5.7) {$\Delta_{i}$};
 \end{tikzpicture}
 \hspace{-0.5cm}
 =
  \sum_{\Delta \in \mathcal{S}^{(t)}} 
D^{\text{bdy}}_\Delta
\hspace{-0.4cm}
\begin{tikzpicture}[baseline=(base), scale = .28]
\coordinate (base) at (0, 0);
\draw (-1,2) -- (0,0) -- (4, 0) -- (5,2);
\draw (-1,-2) -- (0,0);
\draw (4,0) -- (5,-2);
\node [above] at (2, 0){$\Delta$};
\node at (5.5, -2.7) {$\Delta_{j}$};
\node at (5.5, 2.7) {$\Delta_{j}$};
\node at (-1.5, -2.7) {$\Delta_{i}$};
\node at (-1.5, 2.7) {$\Delta_{i}$};
 \end{tikzpicture}
 \label{2pt-bulk}
 \ .
\end{align}
Here, the four-legged diagrams  represent the $s$- and $t$-channel conformal blocks: they  can be computed recursively using the Virasoro algebra  \cite{zam84} and depend only on $c$, the weights of the fields and the cross ratio. On the other hand,  $D^{\text{bulk}}_\Delta$ and $D^{\text{bdy}}_\Delta$ are structure constants that are independent of the cross-ratio $\sigma$: they are the unknowns to be found by solving \eqref{2pt-bulk}. 

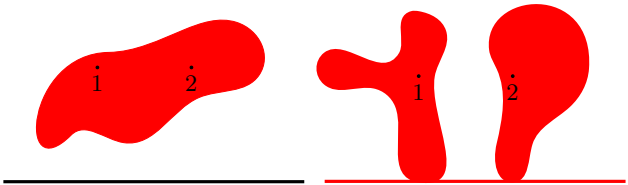
\begin{figure}
\begin{center}
  \begin{tikzpicture}
\coordinate (O) at (5.9,5.5);
\coordinate (P) at (6.6,5.4);
\coordinate (Q) at (7.6,6);
\coordinate (R) at (8.4,6.3);
\coordinate (S) at (8.1,7);
\coordinate (T) at (7.4,6.9);
\coordinate (U) at (6.4,6.6);
\draw[fill = red, draw = red] (O) to [pattern=north east lines, closed, curve through = {(O) (P)  (Q)  (R) (S)  (T) (U)}] 
(O);
 \filldraw[black] (6.25,6.4) circle (0.5pt) node[anchor=north]{1};
 \filldraw[black] (7.5,6.4) circle (0.5pt) node[anchor=north]{2};
    \draw[black, very thick] (5,4.88) -- (9,4.88); 
\end{tikzpicture}
\begin{tikzpicture}
  \draw[thick, fill=red, draw = red]  (6.25,5) to[closed, curve through =
    { (5.95,6.0)(5.65,6.25) (5,6.30)(5,6.70)(6,6.70)(6.15,7.25) (6.25,7.25) (6.6,7) }] (6.45, 6.4);
  \draw[thick, fill=red, draw = red]  (7.5,5) to[closed, curve through =
    { (7.30,5.5)(7.30,6.5) (7.20,6.75)(8.5,6.5)(8.25, 6.0) }] (7.75, 5.5);
 \filldraw[black] (6.25,6.4) circle (0.5pt) node[anchor=north]{1};
 \filldraw[black] (7.5,6.4) circle (0.5pt) node[anchor=north]{2};
    \draw[red, very thick] (5,5) -- (9,5); 
\end{tikzpicture}
         \caption{With free BCs,  two points are connected only if there is a cluster stretching between them: the probability of this decays to zero at large distance. With wired BCs instead, two points even far apart  retain a finite probability of being in the same cluster, since they can do so by both being connected to the wired boundary. }
     \label{Fig}
\end{center}
\end{figure}

We will restrict here to diagonal BCs \cite{djnrs25} where open loop segments cannot end at the boundary. This implies that the bulk $N$-leg operators $V_{(N/2, 0)}$ have vanishing one-point functions (by definition, the loop segments that they insert are not allowed to self-contract). Therefore bulk $N$-leg operators cannot appear in the OPE between any $V_{i}$ and $V_{j}$, so the most general ans\"atze for the spectra of exchanged fields read
\begin{subequations}
\label{ansatz}
\begin{align}
\mathcal{S}^{(s)} &= \{\Delta_{(1,N)}|N \in \mathbb{N}_{>0}\} 
\ , \\
\mathcal{S}^{(t)} &= \{\Delta_{(N,1)} |N \in \mathbb{N}_{>0}\} \ .
\end{align}
\label{spec}
\end{subequations}
With \eqref{ansatz} we used the semi-analytical conformal bootstrap of \cite{djnrs25} to solve \eqref{2pt-bulk} for  $D^{\text{bulk}}_\Delta$ and $D^{\text{bdy}}_\Delta$. We obtained in particular  an  analytic formula for all $D^{\text{bulk}}_\Delta$, which agrees with the numerical results up to 20 digits.  To write below the corresponding expression of
$G_{ij}(\sigma)$, we introduce the function (here $\overset{\scriptscriptstyle 2}{=}$ means that $N$ advances in steps of 2):
\begin{equation} \nonumber
F^{(k)}_{(r_1,s_1)(r_2,s_2)}(\sigma) =   \sum_{N \overset{2}= k}^{\infty} 
C_{(r_1, s_1)(r_2, s_2)}^{(1, N)} R_{(1, N)}
 \hspace{-0.8cm}
 \begin{tikzpicture}[baseline=(base), scale = .28]
\coordinate (base) at (0, 2);
\draw (2, -1) -- (0,0) -- (0, 4) -- (2,5);
\draw (-2, -1) -- (0,0);
\draw (0,4) -- (-2,5);
\node [right] at (0.0,2){$\Delta_{(1, N)}$};
\node at (2.5, -1.7) {$\Delta_{(r_2,s_2)}$};
\node at (2.5, 5.7) {$\Delta_{(r_2,s_2)}$};
\node at (-2.5, -1.7) {$\Delta_{(r_1,s_1)}$};
\node at (-2.5, 5.7) {$\Delta_{(r_1,s_1)}$};
 \end{tikzpicture} \,,
\end{equation}
 where the OPE coefficients 
 were conjectured in \cite{nrj23}: 
\begin{multline} \nonumber
C_{(r_1, s_1)(r_2, s_2)}^{(1, N)} =
   \\
\frac{16P_{(1, N)}\prod_\pm \sin(\pi\beta^{\pm 2})\Gamma_\beta(\pm 2P_{(1, N)})}{\widetilde{\prod} 
\Gamma_\beta\left(\frac{\beta+\beta^{-1}}{2} + \frac{\beta}{2}\left|\epsilon_1r_1+\epsilon_2r_2\right| +\frac{\beta^{-1}}{2}\left(\epsilon_1s_1+\epsilon_2s_2\right) +\epsilon_3P_{(1, N)}\right)}
\ .
\end{multline}
Here $\Gamma_\beta(x)$ is the Barnes double Gamma function, and the product $\widetilde{\prod}$ runs over the eight signs of $\epsilon_i = \pm$.
The coefficients $R_{(1, N)}$ are defined by the one-point functions $\langle V^d_{(1, N)}(z) \rangle=
\frac{R_{(1, N)}}{|z-\bar z|^{2\Delta_{(1, N)}}}$. With the normalization of \cite{djnrs25}, their value is
\begin{equation}
R_{(1, N)} = \sin(2\pi\beta^{-1}P_{(1, N)})
\ .
\end{equation}
\paragraph*{\bf Two-point connectivity $\langle V_{\left (0, 1/2 \right )} V_{\left (0, 1/2 \right )} \rangle$.} 
This two-point function determines the probability that  two bulk points belong to the same FK cluster. There are two solutions to \eqref{2pt-bulk}, compatible with \eqref{spec}, which can be interpreted as describing free and fixed BCs. The latter implies fixing the spin value of every cluster that touches the boundary, so its weight is $Q_1 = 1$. This is equivalent to contracting all the boundary spins into a single one, so we also refer to this as {\em wired} BCs. We find 
\begin{equation}
\label{Gsolution}
G_{{\left (0,\frac{1}{2}\right )} {\left (0, \frac{1}{2}\right )}}
^{\text{free (wired)}}
(\sigma)
=
F^{(1)}_{{\left (0, \frac{1}{2}\right )} {\left (0, \frac{1}{2}\right )}}(\sigma)
\pm
F^{(2)}_{{\left (0, \frac{1}{2},\right )} {\left (0, \frac{1}{2}\right )}}(\sigma)
\ .
\end{equation}
where $+$ is free and $-$ is wired. Note that these expressions, while involving infinite sums, are entirely analytical. The function $G_{{\left (0, 1/2 \right )} {\left (0, 1/2 \right )}}
^\text{wired}
(\sigma)
\overset{\sigma\rightarrow 1}{\sim}
1
$, in agreement with that fact that all points on the boundary are in the same cluster. By contrast, $G_{{\left (0,1/2 \right )} {\left (0, 1/2 \right )}}
^{\text{free}}
(\sigma)
\overset{\sigma\rightarrow 1}{\sim} (1-\sigma)^{\Delta_{(3,1)}}
$, which implies that the leading operator of $\lim_{z \rightarrow \bar z} V_{\left (0, 1/2 \right )}(z)$ is the two-leg boundary operator $v_{(3,1)}$, so clusters touching the boundary are not necessarily connected (see Figure \ref{Fig}).

In order to check these results and provide predictions that can be checked either numerically or against other approaches, we now extract from our formulas some {\sl universal ratios}. 

\paragraph*{\bf Universal ratios.} 

We specialize  to the case of two identical operators with conformal weight $\Delta$, and  denote by $v$ the boundary operator $\lim_{z \rightarrow \bar z} V_{\left (r,s\right )}(z)$, with conformal weight $\Delta_b$. Consider  the two-point function (\ref{12}) in the bulk channel ($\text{Im}(z_i)\to\infty$, or $\sigma \to 0$):
%
\begin{subequations}
\label{Heq}
\begin{equation}
\langle V(z_1) V (z_2) \rangle 
\approx\frac{\lambda}{ |z_1 - z_2|^{4\Delta}} \,, 
\label{Heq1}
\end{equation}
with $\lambda=\lim_{\sigma\to 0} \sigma^{2\Delta}G(\sigma)$ (this normalization constant is denoted $B_P$ in \cite{djnrs25}). Similarly, in the boundary channel ($\text{Im}(z_i) \to 0$, or $\sigma \to 1$) we have
\begin{equation}
\langle V(z_1) V (z_2) \rangle 
\approx\frac{\mu (|z_1-\bar{z}_1||z_2-\bar{z}_2|)^{\Delta_b-2\Delta}}{ |z_1 - z_2|^{2\Delta_b}} \,, 
\label{Heq2}
\end{equation}
\end{subequations}
with $\mu=\lim_{\sigma\to 1} G(\sigma)(1-\sigma)^{2\Delta-\Delta_b}$. 

It is impossible in most cases to adjust the normalization of fields (e.g.\ the Ising spin) on the lattice so that the long-distance limit of their correlation functions  is given exactly by the correlation functions  \eqref{Heq} of the CFT. However, this issue disappears if one forms proper  {\sl ratios} of correlation functions. In the case of interest, this suggests that the ratio $\lambda/\mu$ from the two limits \eqref{Heq} should be measurable from lattice quantities. Note that there is no ambiguity in comparing the $z$ dependencies: since the two correlation functions have the same dimension $(\text{length})^{-4\Delta}$, the renormalization between lattice and continuum distances can be eliminated. 

The ratio  $\lambda/\mu$ can be calculated from our analytical formulas \eqref{2pt-bulk},\eqref{Gsolution}. For wired boundary conditions, we have found a closed-form expression:
\begin{equation}
\left(\frac{\lambda}{\mu}
\right)_{\text{wired}}
= -\frac{\sin(\pi\beta^2)}{\cos(\frac{\pi}{2\beta^{2}})} \,.
\end{equation}
We did  not find a similar expression for free BC, but can obtain arbitrarily accurate numerical values  by computing $D^{\text{bdy}}_{\Delta_{(3,1)}}$ from \eqref{2pt-bulk}. Hence we have clear, simple predictions from our approach to compare with lattice results.

Rather than determine the correlation functions on the half-plane, it is easier technically to carry out calculations on a strip with identical BCs on both sides. Using the standard logarithmic mapping, $w={L\over \pi}\ln z$ with $w=u+iv$ we find in the boundary channel
\begin{multline}
\langle V(w_1)V(w_2)\rangle_{\text{strip}} \approx \\
 \mu\left(\frac{\pi}{L} \right)^{4\Delta} \left[2\sin \left(\tfrac{\pi v_1}{L} \right)2\sin \left( \tfrac{\pi v_2}{L} \right)\right]^{\Delta_b-2\Delta} {\rm e}^{-2\pi u\Delta_b/L} \,, \nonumber
\end{multline}
where $u=u_1-u_2\gg L$.
Obtaining the bulk-channel results from calculations on the strip requires taking both points far from the edge, which would require wider strips than can be reasonable studied. Luckily, we can argue that the asymptotic result (\ref{Heq1}) is exactly the same for calculations on the cylinder instead of the strip. This is because the renormalization between the discrete  operators and those of the CFT is local, and does not depend on BCs.  
Hence we have, still for $u \gg L$,
\begin{eqnarray}
\langle V(w_1)V(w_2)\rangle_{\text{cyl}} &\approx& \lambda\left(\frac{2\pi}{L} \right)^{4\Delta} {\rm e}^{-4\pi u\Delta/L} \,. \label{asympB}
\end{eqnarray}
We can now for simplicity set $v_1=v_2= L/2$ and measure, using transfer matrix calculations, the correlation functions of two operators on the strip and on the cylinder. For $u\gg L$, the exponential contributions arise from the leading eigenvalue of the transfer matrix, while the amplitude is obtained from matrix elements of the lattice operators between the ground state and excited state. Factoring out the resulting $(\pi/L)^{4\Delta}$, we can finally get a lattice estimate of the ratio $\lambda/\mu$.

\paragraph*{\bf Transfer matrix computation.} 

It is convenient to work with the square-lattice FK cluster model \eqref{FKexpansion} on axially oriented strips and cylinders of width $L$ lattice spacings.
In the strip case, vertices of columns 1 and $L$ are considered boundary sites. 
%
%
The spin operator $V_{(0,1/2)}(z)$ corresponds to the order-parameter operator ${\cal O}_a(z_i) = \delta_{\sigma_i,a} - \tfrac{1}{Q}$ in the
Potts model, where $a \in \{1,\ldots,Q\}$ is some fixed color.

To access $\langle V_{(0,1/2)} V_{(0,1/2)} \rangle$ on the lattice, we first remark that
with free or periodic BCs $\langle \sum_a {\cal O}_a(z_1) {\cal O}_a(z_2) \rangle$ is proportional, after FK expansion, to the probability
$P(z_1 \sim z_2)$ that $z_1$ and $z_2$ belong to the same FK cluster. With fixed BCs on the strip, the FK expansion of
$\sum_a {\cal O}_a(z_1) {\cal O}_a(z_2)$ is more involved: a cluster has weight $0$ if it contains precisely one point among $z_1$ or $z_2$ and no boundary site,
and otherwise weight $Q$ (resp.\ $1$) if it contains no (resp.\ at least one) boundary site. This is simpler stated if we consider all boundary sites to
be implicitly connected (wired BCs), in which case $\langle \sum_a {\cal O}_a(z_1) {\cal O}_a(z_2) \rangle$ is proportional to $P(z_1 \sim z_2)$
as before, and the cluster weights are as in \eqref{FKexpansion} with $Q_1 = 1$. We notice also that in the original Potts model, formulated in terms
of $Q$-component spins, $\langle \sum_a {\cal O}_a(z_1) {\cal O}_a(z_2) \rangle$ is equal to $\langle \sigma_1 \sigma_2 \rangle$ for the Ising model ($\sigma_i = \pm 1$)
and to $\langle \sigma_1 \sigma_2^* \rangle$ for the 3-state Potts model ($\sigma_i = 1, \omega, \omega^2$, with $\omega = \exp(2 \pi i/3)$), exactly on any finite lattice.

\begin{figure}
\begin{center}
\includegraphics[width=0.23\textwidth]{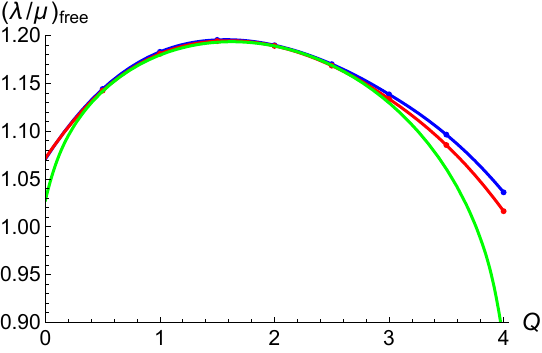} \
\includegraphics[width=0.23\textwidth]{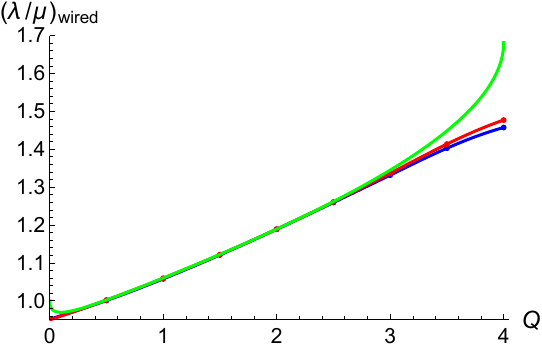}
\end{center}
\caption{Ratio $\lambda / \mu$ for the $Q$-state FK model. The BCs are free (left panel) and wired (right panel).
The transfer-matrix results (blue curves using $L=5,7,9$; red curves using $L=5,7,9,11$) converge well to the bootstrap results (in green).
}
\label{fig:ratio}
\end{figure}

The computation of $\langle \sum_a {\cal O}_a(z_1) {\cal O}_a(z_2) \rangle$ in the FK expansion proceeds by the transfer-matrix techniques of \cite[Appendix B]{js18}.
States are set partitions of the vertex set $\{1,\ldots,L\}$ of a row, in which some of the parts can be labelled as containing at least one boundary site, or 
one point among $z_1$ or $z_2$ and no boundary site, so as to correctly produce the weights given above. The operators ${\cal O}_a(z_i)$ are inserted
in the middle column $(L+1)/2$, using only odd $L$. In practice we use $L=5,7,9,11$. We extra\-polate to the $L \to \infty$ limit by fitting the lattice estimates to polynomials
in $1/L$, using the first three or all four available sizes. The agreement with the bootstrap results is excellent, except near $Q=4$ where we encounter
the well-known \cite{Cardy86_log} logarithmic corrections to scaling.

\paragraph*{\bf Conclusion and outlook.} 

In this Letter, we proposed a conformal bootstrap approach for computing two-point functions of bulk fields in critical loop models in a boundary geometry. The two-point functions were assumed to be solutions of the crossing-symmetry equation that are consistent with the physical BCs. We  applied our approach to the two-point connectivity $\langle V_{\left (0, 1/2 \right )} V_{\left (0, 1/2 \right )} \rangle$, and argued that the amplitude ratio $\lambda / \mu$ is insensitive to the different field normalizations in the lattice model and the CFT limit. The bootstrap prediction for $\lambda / \mu$ was indeed found in excellent agreement with transfer-matrix computations, in the limit $L \to \infty$ of large lattices.

Earlier results on correlation functions for boundary loop models include Schramm's left-passage probability \cite{schramm01} and Cardy's crossing probability \cite{car91}. Both are linear combinations (satisfying BCs) of two conformal blocks, that are hypergeometric functions. By contrast, our two-point connectivity \eqref{Gsolution} is a linear combination of two $F$-functions, each of which is an {\em infinite} linear combination of conformal blocks.

We expect that our approach can also be used with other bulk two-point functions. For instance, consider the two-point function $\langle V_{\left (N/2, 0\right )} V_{\left (M/2, 0\right )} \rangle $ with free BC. For this case, with the ansatz \eqref{spec}, there is only one solution to the crossing  equation \eqref{2pt-bulk}:
\begin{equation}
G_{{\left (\frac{N}{2}, 0\right )} {\left (\frac{M}{2}, 0\right )}}(\sigma)
=
F^{(1)}_{{\left (\frac{N}{2}, 0\right )} {\left (\frac{M}{2}, 0\right )}}(\sigma)
\ .
\label{fn1}
\end{equation}
The function in \eqref{fn1} is only non-zero for $N+M$ even, and this implies that the loop segments inserted by
$V_{\left (N/2, 0\right )}$ and $V_{\left (M/2, 0\right )}$ cannot end on the boundary, in agreement with our interpretation of the free BC. 

Our approach can also determine correlation functions for other, maybe less familiar BCs such as Dirichlet  --- where lines can end on the boundary --- for $O(n)$, or the ``new'' BCs for the Potts and FK models \cite{Ian_Affleck_1998}.

To conclude, we note that the important question of {\sl classification} of all possible conformal boundary conditions in critical loop models  remains totally open.

\subsection*{Acknowledgements}
We are very grateful to Sylvain Ribault for his contribution in \cite{djnrs25} that led to results in this letter.
RN thanks Federico Camia, Xin Sun, and Baojun Wu for useful discussion. MD thanks Paul Roux for insightful discussions.
This work is supported by the French Agence Nationale de la Recherche (ANR) under grant ANR-21-CE40-0003 (project CONFICA).

\bibliography{../../inputs/992.bib}


\end{document}